\documentclass[journal]{IEEEtran}
\usepackage{geometry}
 \usepackage{amsmath}
 \usepackage{subfigure}
 \usepackage{graphicx,graphics,color,psfrag}
 \usepackage{cite,balance}
 \usepackage{amsmath,amssymb,amsfonts}
 \usepackage{caption}
 \usepackage{amsmath}
 \usepackage{amsthm}
\usepackage{textcomp}
 \allowdisplaybreaks
 \usepackage{algorithm}
  \usepackage{algorithmic}
 \usepackage{accents}
 \usepackage{amsthm}
 \usepackage{bm}
  \usepackage{url}
 \usepackage[english]{babel}
 \usepackage{multirow}
 \usepackage{enumerate}
 \usepackage{cases}
 \usepackage{stfloats}
 \usepackage{dsfont}
 \usepackage{color,soul}
 \usepackage{amsfonts}
 \usepackage{cite,graphicx}
 \usepackage{subfigure}
 \usepackage{fancyhdr}
 \usepackage{hhline}
 \usepackage{graphicx,graphics}
 \usepackage{array,color}
  \usepackage{bm}

\newtheorem{lemma}{\emph{\underline{Lemma}}}

\newtheorem{proposition}{\emph{\underline{Proposition}}}

\def\l{\left}
\def\r{\right}
\def\({\left(}
\def\){\right)}

\def\b0{{\mathbf{0}}}

\newcommand{\diag}{\mathrm{diag}}

\newcommand{\nn}{\nonumber}

\renewcommand{\baselinestretch}{0.97}
\begin{document}
\title{Delay-Optimal Scheduling for IRS-Aided \\ Mobile Edge Computing}
\author{{Fasheng Zhou,~\emph{Member,~IEEE}, Changsheng You,~\emph{Member,~IEEE},
	 and Rui Zhang,~\emph{Fellow,~IEEE}}  \thanks{\noindent F. Zhou is with Guangzhou University, Guangzhou, China (Email: zhoufs@gzhu.edu.cn). C. You and R. Zhang are with National University of Singapore, Singapore (Email: \{eleyouc, elezhang\}@nus.edu.sg).}
	
}
\maketitle 
\begin{abstract}
We consider an intelligent reflecting surface (IRS)-aided mobile edge computing (MEC) system, where an IRS is deployed to assist computation offloading from two users to an access point connected with an edge cloud. For the IRS-aided data transmission, in contrast to the conventional non-orthogonal multiple-access (NOMA) and  time-division multiple-access (TDMA), we propose a new flexible \emph{time-sharing NOMA} scheme that allows users to flexibly divide their data into two parts transmitted via NOMA and TDMA, respectively, thus encapsulating both conventional NOMA and TDMA as special cases. We formulate an optimization problem to minimize the sum delay of the two users by designing the IRS passive reflection and users' computation-offloading scheduling under the IRS discrete-phase constraint. Although this problem is non-convex, we obtain its optimal solution for both the cases of infinite and finite cloud computing capacities. Furthermore, we  show that NOMA and TDMA based transmissions are preferred in different scenarios, depending on the users' cloud-computing time as well as the rate discrepancy between NOMA and TDMA with IRS. 

\end{abstract}
\IEEEpeerreviewmaketitle


%%%%%%%%%%%%%%%%%%%%%%%%%%%%%%%%%%%%%%%%%%%%%%%%%%%%%%%%%%%%%%%%%%%%%%%%%%%%%%%%%%%%%%%%%%%%%%
\section{Introduction}
%\IEEEPARstart{O}
Intelligent reflecting surface (IRS) has emerged as a promising technology to enhance the spectral efficiency of wireless communication systems cost-effectively, by leveraging its massive reflecting elements to engineer the radio propagation environment in favor of signal transmission \cite{wu2020,Basar}. Moreover, compared to the conventional  active relay, IRS reflects signals without any active component and operates in a full-duplex and noise-free manner. These appealing advantages have motivated active research on IRS recently to integrate IRS into conventional wireless systems/networks for enhancing their communication performance, leading to various new applications, such as  IRS-aided non-orthogonal multiple-access (NOMA) \cite{bxiong,gcirsnoma}, IRS-aided physical-layer security \cite{Guan_2020}, IRS-aided mobile edge computing (MEC) \cite{irsmec1,cao2019}, etc. 

In particular, for the IRS-aided MEC system, the existing works (e.g., \cite{irsmec1,cao2019}) have shown that properly deploying an IRS near the offloading users and designing its passive beamforming efficiently can help reshape the computation-load distributions among the users and enhance the user-cloud communication rates, thus greatly reducing their computation-offloading delays especially when the user-cloud direct links are blocked. However, these works  simply assumed either simultaneous or sequential data transmission for the users, thus neglecting the effects of IRS on designing users' computation-offloading scheduling under different multiple-access schemes. To be specific, for a two-user MEC system without IRS, it has been shown in  \cite{shenmin} that if assuming negligible cloud-computing time, the NOMA-based data transmission achieves smaller sum (data-transmission) delay than that based on orthogonal multiple-access (OMA), e.g., time-division multiple-access (TDMA). This result, however, may not still hold for the IRS-aided MEC system with the practically non-negligible cloud-computing time  due to the following reasons. First, different from the conventional NOMA-based transmission for which the cloud can compute tasks only after receiving all the data, the TDMA-based transmission allows the cloud to compute the task of the firstly-scheduled user when the second user transmits its data, thus leading to the \emph{transmission-computing multiplexing gain} for reducing the sum delay. Second, in terms of data transmission, IRS reflection design needs to balance the rate improvement among the users when employing NOMA, while for TDMA, the IRS is capable of optimally combining signals of individual user separately by leveraging its time-selectivity. Thus, the delay-optimal scheduling and multiple-access scheme for the IRS-aided MEC system are still unknown in general, to the authors' best knowledge.

\begin{figure}[t]
  \centering
  \includegraphics[width=2.5in]{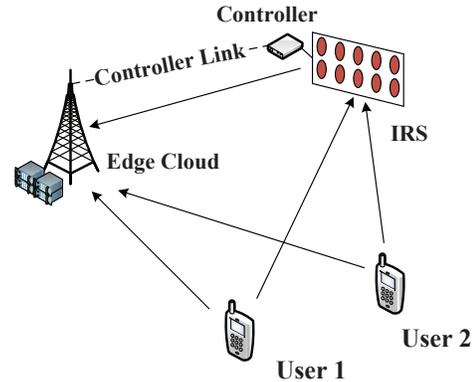}
  \caption{An IRS-aided MEC system.}
  \label{model}
  \end{figure}

\begin{figure}[t]
 \centering
 \includegraphics[width=3.5in]{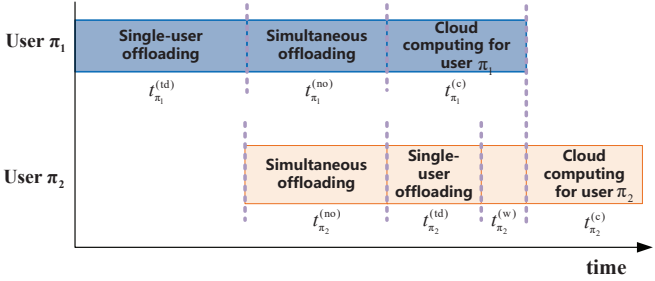}
 \caption{Illustration of the proposed time-sharing NOMA scheme for computation offloading.}
 \label{tmConsmptn}
 \end{figure}

To address the above issue, we consider an IRS-aided MEC system as shown in Fig.~\ref{model}, where an IRS is deployed near two users \footnote{The results in this paper for the two-user NOMA can be extended to the more general case with a larger number of users by integrating the two-user power control with efficient user pairing \cite{userpairing}.} to assist their computation offloading to an access point (AP) connected with an edge cloud. We propose a flexible  time-sharing NOMA scheme for data transmission as illustrated in Fig.~\ref{tmConsmptn} that allows users to flexibly divide their data into two parts transmitted via NOMA and TDMA, respectively, thus encapsulating both conventional NOMA and TDMA as special cases. An optimization problem is formulated to minimize the sum delay of the two users by jointly designing users' scheduling order, time division, IRS discrete phase shifts, users' transmit power as well as signal decoding order when employing NOMA.  Although this problem is non-convex, we obtain its optimal solution for both the cases of infinite and finite cloud computing capacities. Furthermore, we show that  TDMA is better than NOMA for the sum delay minimization when the users' cloud-computing time is sufficiently long and/or the rate improvement of TDMA against NOMA, both  with IRS, is sufficiently large; whereas the reverse is true otherwise.

\section{System Model and Problem Formulation}\label{systemmodel}
\subsection{System Model} 
We consider an IRS-aided  MEC system as shown in Fig.~\ref{model}, where an IRS is deployed to assist computation offloading from two single-antenna users to a single-antenna AP, which is the gateway of an edge cloud. The IRS has $N$ reflecting subsurfaces, each consisting of $M$ elements. We denote by $\boldsymbol{\Theta}\triangleq\diag(e^{\jmath\omega_1},  \cdots,  e^{\jmath\omega_{N}})\in\mathbb{C}^{N\times N}$ the IRS diagonal reflection matrix, where the reflection amplitude of each subsurface is set to be one  for simplicity and $\omega_n$, $n\in\mathcal{N}\triangleq\{1,\cdots, N\}$ denotes the phase-shift of subsurface $n$. Let $Q$ denote the number of practical discrete phase-shift levels. Then we have $\omega_n\in\mathcal{F}\triangleq\{0, \Delta\omega, \cdots, (Q-1)\Delta\omega\}$, where  $\Delta\omega=2\pi/Q$. Moreover, we denote by $h_{{\rm d},k}\in\mathbb{C}$, $\boldsymbol{h}_{{\rm r},k}\in\mathbb{C}^{N\times 1}$, and $\boldsymbol{h}^H \in\mathbb{C}^{1\times N}$ the baseband channels from user $k$ to  AP, user $k$ to IRS, and IRS to AP, respectively, with $k\in\mathcal{K}\triangleq\{1,2\}$.\footnote{We assume that the channel state information (CSI) of all involved channels is available at the AP, which can be practically obtained by various channel estimation methods (see, e.g., \cite{cs_jsca_irs}).} As such, the effective channel from user $k$ to the AP is given by 
 \begin{equation}
 g_k=\boldsymbol{h}^H\boldsymbol{\Theta}\boldsymbol{h}_{{\rm r},k}+h_{{\rm d},k}\triangleq \boldsymbol{g}_{k}^H \boldsymbol{\theta}+h_{{\rm d},k},
 \end{equation}
 where $\boldsymbol{g}_{k}^H\triangleq \boldsymbol{h}^H\diag(\boldsymbol{h}_{{\rm r},k})$ and $\boldsymbol{\theta}\triangleq[e^{\jmath\omega_1},  \cdots,  e^{\jmath\omega_{N}}]^T$.

For each user $k\in\mathcal{K}$, let $L_k$ in bits denote the data size of its task input-data required to be offloaded\footnote{The proposed scheme assuming full offloading can be extended to the case with partial offloading by further optimizing the data partitioning of the two users for local computing and computation offloading separately \cite{MECsurvey}.} and $C_k$ in cycles/bit denote the number of  required  CPU cycles for computing $1$-bit data.  
To reduce the sum offloading delay, we propose a new \emph{time-sharing NOMA} scheme for the IRS-aided computation offloading as illustrated in Fig.~\ref{tmConsmptn}. Specifically, we denote by $\boldsymbol \pi\triangleq\{\pi_1,\pi_2\}$ the offloading scheduling order for the two users with $\pi_m\in\mathcal{K}$ for $m=1,2$, under which user $\pi_1$ is scheduled before user $\pi_2$. For data transmission, user $\pi_1$ firstly transmits part of its data alone over a time duration of $t_{\pi_1}^{(\rm td)}$ and then shares a common transmission time of duration $t^{(\rm no)}$ with user $\pi_2$  by employing  NOMA, after which user $\pi_2$ sends its remaining input data within a time duration of $t_{\pi_2}^{(\rm td)}$. Note that the proposed flexible time-sharing NOMA scheme reduces to the conventional NOMA  and TDMA when we have $t_{\pi_1}^{(\rm td)}=t_{\pi_2}^{(\rm td)}=0$ and  $t^{(\rm no)}=0$, respectively.
For task computing, we assume that the cloud has a single-core CPU with $F$  in Hertz (Hz) denoting its CPU-cycle frequency and it computes the tasks for the users in a first-come-first-serve manner. 
As such, the cloud firstly serves user $\pi_1$ at the time instant $t_{\pi_1}^{(\rm td)}+t^{(\rm no)}$ and then serves user $\pi_2$ after finishing the task computing for user $\pi_1$ and receiving all input data from user $\pi_2$. Let $t_{\pi_m}^{(\rm c)}\triangleq\frac{C_{\pi_m} L_{\pi_m}}{F}$ denote the task-computing duration for user $\pi_m\in\mathcal{K}$. Then, the (task-computing) waiting time for user $\pi_2$ after offloading all its data is given by  $t_{\pi_2}^{(\rm w)}=\max(0,t_{\pi_1}^{(\rm c)}-t_{\pi_2}^{(\rm td)})$. Similar to \cite{cs_mec_twc}, we assume small data sizes for the computation results of the two users (e.g., face recognition, object detection in video, and online chess game) and thus the result-feedback delay is negligible. Based on the above, the computation offloading delay for each user $\pi_m$ is given by
\begin{align}
T_{\pi_1}&=t_{\pi_1}^{(\rm td)}+t^{(\rm no)}+t_{\pi_1}^{(\rm c)},\label{T_pi1}\\
T_{\pi_2}&=t_{\pi_1}^{(\rm td)}+t^{(\rm no)}+t_{\pi_2}^{(\rm td)}+t_{\pi_2}^{(\rm w)}+t_{\pi_2}^{(\rm c)},\label{T_pi2}
\end{align}
and thus the sum delay of the two users is $T_{\pi_2}$.

Next, we model the IRS-aided data transmission as follows. For the separate transmission of each user $\pi_m$, it can be easily shown that the optimal IRS passive reflection for maximizing its achievable rate is given by $\angle[\boldsymbol{\theta}_{\pi_m}^{(\rm td)}]_n = {\rm quan}(\angle h_{{\rm d}, \pi_m}\!-\!\angle[\boldsymbol{g}_{\pi_m}]_n)$, $n\in\mathcal{N}$, where ${\rm quan}$ denotes the nearest-phase-quantization operation proposed in \cite{cs_jsca_irs}; thus the corresponding achievable rate of  user $\pi_m$ in bits/second (bps) is given by 
\begin{equation}
r_{\pi_m}^{(\rm td)}=B {\rm log}_2\l(1+\frac{P_{\pi_m} f_{\pi_m}(\boldsymbol{\theta}_{\pi_m}^{(\rm td)})}{\sigma^2}\r),
\end{equation}
where $f_{\pi_m}(\boldsymbol{\theta}_{\pi_m}^{(\rm td)})\triangleq |\boldsymbol{g}_{{\pi_m}}^H \boldsymbol{\theta}_{\pi_m}^{(\rm td)}+h_{{\rm d},{\pi_m}}|^2$ denotes the effective channel power gain for user $\pi_m$, $P_{\pi_m}$ is the maximum transmit power of user $\pi_m$, $B$ denotes the system bandwidth in Hz, and $\sigma^2$ denotes the received noise power at the AP. On the other hand, for the NOMA-based  transmission, as  users' effective channel gains are determined by the IRS passive reflection, we have $2!=2$ possible decoding orders for the users at the AP. For example, if user $\pi_1$ is decoded before user $\pi_2$,  the achievable rates of the two users are given by
\begin{align}
\label{nomarate1}
r_{\pi_1}^{(\rm no)}&=B {\rm log}_2\l(1+\frac{p_{\pi_1} f_{\pi_1}(\boldsymbol{\theta}^{(\rm no)})}{p_{\pi_2} f_{\pi_2}(\boldsymbol{\theta}^{(\rm no)})+\sigma^2}\r),\\ \label{nomarate2}
r_{\pi_2}^{(\rm no)}&=B {\rm log}_2\l(1+\frac{p_{\pi_2} f_{\pi_2}(\boldsymbol{\theta}^{(\rm no)})}{\sigma^2}\r),
\end{align}
where $\boldsymbol{\theta}^{(\rm no)}$ denotes the IRS passive reflection for  NOMA, and $0\le p_{\pi_1}\le P_{\pi_1}$ and $0\le p_{\pi_2}\le P_{\pi_2}$ denote the transmit power of users $\pi_1$ and $\pi_2$, respectively. For the other case with user $\pi_2$ decoded before user $\pi_1$, we denote the users' achievable rates by $\tilde{r}_{\pi_1}^{(\rm no)}$ and $\tilde{r}_{\pi_2}^{(\rm no)}$, respectively, which can be similarly defined as \eqref{nomarate1} and \eqref{nomarate2} and thus are omitted.

\subsection{Problem Formulation}
Our objective is to minimize the sum delay of the two users, $T_{\pi_2}$, by jointly optimizing users' scheduling order ${\boldsymbol \pi}$,  time division for data transmission $\boldsymbol{t}\triangleq \{t_{\pi_1}^{(\rm td)}, t_{\pi_2}^{(\rm td)}, t^{(\rm no)}\}$, IRS passive reflection $\boldsymbol{\theta}^{(\rm no)}$, users' transmit power $\boldsymbol{p}\triangleq\{p_{\pi_1},p_{\pi_2}\}$, as well as users' information decoding order, under the constraints on users' maximum transmit power, IRS discrete phase shifts, and the users' individual offloaded data size. This problem can be decomposed into two subproblems associated with two different information decoding orders. For brevity, we formulate one of the subproblems as follows corresponding to the case with user ${\pi_1}$ decoded first, while the other case can be similarly formulated and solved, thus is omitted.
\begin{subequations}
\begin{align}
%\text{P1}
({\bf P1}): ~ \mathop {\min }\limits_{{\boldsymbol \pi}, {\boldsymbol t}, \boldsymbol{\theta}^{(\rm no)}, \boldsymbol{p}} & ~T_{\pi_2} \nn\\ 
 ~{\rm s.t.}~~~ & t_{\pi_1}^{(\rm td)} r_{\pi_1}^{(\rm td)} + t^{(\rm no)} r_{\pi_1}^{(\rm no)} = L_{\pi_1},\\ 
            & t_{\pi_2}^{(\rm td)} r_{\pi_2}^{(\rm td)} + t^{(\rm no)} r_{\pi_2}^{(\rm no)} = L_{\pi_2},  \\ 
            & t_{\pi_1}^{(\rm td)}\ge 0, t_{\pi_2}^{(\rm td)}\ge 0, t^{(\rm no)}\ge 0,\\
            & \angle[\boldsymbol{\theta}^{(\rm no)}]_n\in\mathcal{F}, ~~~\forall n\in\mathcal{N},\\
            & 0\le p_{\pi_1}\le P_{\pi_1}, 0\le p_{\pi_2}\le P_{\pi_2},        
 \end{align}
 \end{subequations}
 where $T_{\pi_2}$ is given in \eqref{T_pi2}, and $r_{\pi_m}^{(\rm no)}$ is jointly determined by $\boldsymbol{\theta}^{(\rm no)}$ and $\boldsymbol{p}$ as given in \eqref{nomarate1} and \eqref{nomarate2}.

It can be easily verified that problem (P$1$) is a non-convex optimization problem due to the  discrete-phase constraint as well as the non-convex function for  $r_{\pi_m}^{(\rm no)}$. Nevertheless, this problem is solved optimally in the next section.

\section{Proposed Solution}
In this section, we first solve problem (P$1$) for the case of infinite cloud (computing) capacity and then extend the solution to the general case with finite cloud capacity. 

\subsection{Infinite Cloud Capacity}\label{Sec:Infinite} 
Given infinite cloud capacity which corresponds to the practical scenario with sufficiently high cloud computational capacity, the task-cloud-computing time for both users as well as the waiting time for user $\pi_2$  are negligible. Then, it can be easily shown that both scheduling orders are optimal to problem (P$1$) and thus we simply consider the case with $\{\pi_1=1, \pi_2=2\}$, without loss of optimality. As such, the sum delay reduces to  
%$T_{1}=t_{1}^{(\rm td)}+t^{(\rm no)}$ and 
$T_{2}=t_{1}^{(\rm td)}+t^{(\rm no)}+t_{2}^{(\rm td)}$.  Although much simplified, problem (P$1$) is still non-convex and thus difficult to solve. To tackle this problem, we first consider the optimization problem of transmission time division $\boldsymbol{t}$ with fixed 
$\{\boldsymbol{\theta}^{(\rm no)}, \boldsymbol{p}\}$ and then find the optimal $\{\boldsymbol{\theta}^{(\rm no)}, \boldsymbol{p}\}$ to solve problem (P$1$). 

First, given any feasible $\{\boldsymbol{\theta}^{(\rm no)}, \boldsymbol{p}\}$ (and hence fixed $\{r_{1}^{(\rm no)},r_{2}^{(\rm no)}\}$),  problem (P$1$) can be transformed into the following equivalent form after some algebraic manipulations.
\begin{subequations}
\begin{align}
%\label{P2}
\!\!\!\!({\bf P2}): 	 \mathop {\min }\limits_{{\boldsymbol t}} ~~&    t_{1}^{(\rm td)}+t^{(\rm no)}+t_{2}^{(\rm td)} \nn\\ 
 {\rm s.t.}~~ &t_{1}^{(\rm td)} r_{1}^{(\rm td)} + t^{(\rm no)} r_{1}^{(\rm no)} = L_1,\label{Eq:p2:1}\\ 
            & t_{2}^{(\rm td)} r_{2}^{(\rm td)} + t^{(\rm no)} r_{2}^{(\rm no)} = L_2,\label{Eq:p2:2}  \\ 
&  0 \le t_{1}^{(\rm td)}  \le L_1/r_{1}^{(\rm td)},  0 \le t_{2}^{(\rm td)} \le L_2/r_{2}^{(\rm td)}, \label{Eq:p2:2} \\ 
              & 0 \le t^{(\rm no)} \le \min\{L_1/r_{1}^{(\rm no)},L_2/r_{1}^{(\rm no)}\}. \label{Eq:p2:4}
\end{align}
\end{subequations}
Problem (P$2$) is a linear programming (LP) and thus can be easily solved. By defining $\lambda(\boldsymbol{\theta}^{(\rm no)}, \boldsymbol{p})\triangleq r_{1}^{(\rm no)}/r_{1}^{(\rm td)}  + r_{2}^{(\rm no)}/r_{2}^{(\rm td)}-1$ as the \emph{(IRS-dependent) NOMA priority}, the optimal solution to problem (P2) is obtained as follows.

\begin{proposition}\label{prop:infinite} \emph{With $\lambda(\boldsymbol{\theta}^{(\rm no)}, \boldsymbol{p})$, the optimal solution to problem (P$2$) is given by
\begin{itemize}
\item[1)] If $\lambda(\boldsymbol{\theta}^{(\rm no)}, \boldsymbol{p})\ge 0$, we have
\begin{align}
\label{t0opt}
t^{(\rm no)*} &= \min\{L_1/r_{1}^{(\rm no)},L_2/r_{2}^{(\rm no)}\},\\
\label{tkopt}
t_{k}^{(\rm td)*}& =\frac{L_k-t^{(\rm no)*} r_k^{(\rm no)}}{r_k^{(\rm td)}}, \forall k\in\mathcal{K}.
\end{align}
\item[2)] Otherwise, if $\lambda(\boldsymbol{\theta}^{(\rm no)}, \boldsymbol{p})<0$, we have 
$$t^{(\rm no)*}=0, ~t_{k}^{(\rm td)*}=L_k/r_k^{(\rm td)}, \forall k\in\mathcal{K}.$$
\end{itemize}
Moreover, the minimum sum delay is given by 
\begin{align}
\label{sum1}
\!\!T_2^*&=L_1/r_1^{(\rm td)}+L_2/r_2^{(\rm td)}-\lambda(\boldsymbol{\theta}^{(\rm no)}, \boldsymbol{p}) t^{(\rm no)*}\nn\\
&=L_1/r_1^{(\rm td)}+L_2/r_2^{(\rm td)}\nn\\
&~~~-\max\{\lambda(\boldsymbol{\theta}^{(\rm no)}, \boldsymbol{p}),0\} \min\{L_1/r_{1}^{(\rm no)},L_2/r_{2}^{(\rm no)}\}.
\end{align}
 }
 \end{proposition} 
 
Proposition~\ref{prop:infinite} shows that given the IRS passive reflection and users' transmit power, the optimal time division for the proposed time-sharing NOMA scheme has a \emph{threshold-based} structure.
Specifically, if the NOMA priority $\lambda(\boldsymbol{\theta}^{(\rm no)}, \boldsymbol{p})$ is non-negative, the two users should try to transmit their data based on NOMA and at least one of them offloads all  its data via NOMA only; otherwise, the two users should separately transmit their data based on TDMA since the NOMA-based transmission suffers rate-performance loss. Moreover, it is worth noting that different from the conventional uplink system without IRS for which the NOMA priority is always positive as NOMA is superior to TDMA in terms of achievable rate, the current NOMA priority, however, critically depends on the IRS passive reflection since it needs to be set identical for the two users in the case of NOMA and thus may not perfectly align signals for both users at the same time, thus resulting in certain rate degradation.

With Proposition~\ref{prop:infinite}, it can be shown that problem (P$1$) is equivalent to the following problem for optimizing the discrete phase shifts and users' transmit power.
\begin{subequations}
\begin{align}
\label{P3}
\!\!\!\!({\bf P3}): 	 \mathop {\max }\limits_{\boldsymbol{\theta}^{(\rm no)}, \boldsymbol{p}} ~&   
% \max\l\{
 \l(\frac{r_{1}^{(\rm no)}}{r_{1}^{(\rm td)}}  + \frac{r_{2}^{(\rm no)}}{r_{2}^{(\rm td)}}-1\r)
% ,0\r\}
 \min\l\{\frac{L_1}{r_{1}^{(\rm no)}},\frac{L_2}{r_{2}^{(\rm no)}}\r\} \nn\\ 
 {\rm s.t.}~~ & \angle[\boldsymbol{\theta}^{(\rm no)}]_n\in\mathcal{F}, ~~~\forall n\in\mathcal{N},\\
            & 0\le p_{1}\le P_{1}, 0\le p_{2}\le P_{2}.
\end{align}
\end{subequations}
Problem (P$3$) is a non-convex optimization problem due to the non-concave objective function and non-convex constraints. To tackle this difficulty, we first present a useful lemma below.
\begin{lemma}\label{Lem:p1}\emph{Given any feasible $\boldsymbol{\theta}^{(\rm no)}$, 
the optimal $\boldsymbol{p}$ to problem (P$3$) is given by $p_1^*=P_{1}$ and $p_2^*=P_{2}$.
}
\end{lemma}
\noindent\emph{Sketch of proof:}
First, it is noted that the increase of $p_1$ leads to an increasing $r_{1}^{(\rm no)}$ without affecting $r_{2}^{(\rm no)}$. Then, by showing that given $r_{2}^{(\rm no)}$, the objective value of problem (P$3$), denoted by $\Gamma$,  monotonically increases with $r_{1}^{(\rm no)}$, 
we can conclude that given 
$\{\boldsymbol{\theta}^{(\rm no)},p_2\}$, $\Gamma$ monotonically increases with $p_1$. Next, since the increase of $p_2$ leads to an increasing $r_{2}^{(\rm no)}$ but a decreasing $r_{1}^{(\rm no)}$, we consider the following two cases. First, if $L_1/r_{1}^{(\rm no)}\!<\!L_2/r_{2}^{(\rm no)}$, it can be shown that the derivative of $\Gamma$ with respect to $p_2$ is always positive and thus $p_2^*=P_2$. For the other case of $L_1/r_{1}^{(\rm no)}\ge L_2/r_{2}^{(\rm no)}$, user $1$ still needs to transmit data after the NOMA time duration $t^{(\rm no)*}$. It can be shown that increasing $p_2$  results in less  transmission time for user $2$ (i.e., $t^{(\rm no)} = L_2/r_{2}^{(\rm no)}$), due to its higher transmission rate $r_{2}^{(\rm no)}$.
This in turn reduces the total transmission delay of user $1$, since more data is transmitted during its separate transmission with a higher achievable rate than that during $t^{(\rm no)}$.
Combining the above results leads to Lemma~\ref{Lem:p1}.
\hfill $\Box$

Using Lemma~\ref{Lem:p1}, the optimal solution to problem (P$3$) can be numerically obtained by searching over all possible combinations of IRS discrete phase shifts with $
\{p_1=P_{1}, p_2=P_{2}\}$  and then choosing the one that achieves the maximum value of problem (P$3$).
To reduce the computational complexity of the exhaustive search for $\boldsymbol{\theta}^{(\rm no)}$ when $Q$ is large,  we further apply a low-complexity algorithm proposed in \cite{bxiong} to sub-optimally solve problem (P$3$) by defining $\boldsymbol{\theta}^{(\rm no)}$ as  ${\boldsymbol{\theta}}^{(\rm no)}\triangleq\eta\boldsymbol{\theta}_{1}^{(\rm td)}+(1-\eta)\boldsymbol{\theta}_{2}^{(\rm td)}$ with $0\le\eta\le 1$ and thus reducing the exhaustive search for ${\boldsymbol{\theta}}^{(\rm no)}$ to a one-dimensional search for $\eta$ followed by the discrete-phase quantization. 

\subsection{Finite Cloud  Capacity}
Next, we consider the general case with finite cloud capacity, for which the cloud-computing time for the two users is non-negligible and thus affects the optimal scheduling order. In general, problem (P1) can be decomposed into
two subproblems associated two different scheduling orders. For brevity, we solve the subproblem associated with $\{\pi_1 = 1, \pi_2 = 2\}$ in this subsection, while the other case can be similarly solved. 

Following the similar procedures in Section~\ref{Sec:Infinite}, we first fix $\{\boldsymbol{\theta}^{(\rm no)}, \boldsymbol{p}\}$ and thus reduce problem (P$1$) as
\begin{subequations}
\label{P4}
\begin{align}
\!\!\!\!({\bf P4}): 	 \mathop {\min }\limits_{{\boldsymbol t}} ~~&    t_{1}^{(\rm td)}+t^{(\rm no)}+t_{2}^{(\rm td)}+\max(0,t_{1}^{(\rm c)}-t_{2}^{(\rm td)})+t_{2}^{(\rm c)} \nn \\ 
 {\rm s.t.}~~ & \eqref{Eq:p2:1}-\eqref{Eq:p2:4}.\nn
\end{align}
\end{subequations}
Problem (P$4$) is a convex optimization problem,
whose optimal solution is obtained in closed-form as follows.

\begin{proposition}\label{prop2}\emph{The optimal solution to problem (P$4$) is 
\begin{itemize}
\item[1)] If i) $t_{1}^{(\rm c)}\ge L_2/r_2^{(\rm td)}$ or ii) $t_{1}^{(\rm c)}< L_2/r_2^{(\rm td)}$ and $\lambda(\boldsymbol{\theta}^{(\rm no)}, \boldsymbol{p}) < 0$,  we have
\begin{align}
t^{(\rm no)*}&=0, t_{k}^{(\rm td)*}=L_k/r_k^{(\rm td)}, \forall k\in\mathcal{K},\label{p4case1}
\end{align}
which leads to $T_2^*=L_1/r_1^{(\rm td)} \!+\! \max\{t_{1}^{(\rm c)}, L_2/r_2^{(\rm td)}\} \!+\! t_{2}^{(\rm c)}$.
{\item[2)] Otherwise, if $t_{1}^{(\rm c)}< L_2/r_2^{(\rm td)}$ and $\lambda(\boldsymbol{\theta}^{(\rm no)}, \boldsymbol{p}) \ge 0$,
  we have
\begin{align}
t_{2}^{(\rm td)*}  &= \max\l\{\frac{L_2-(L_1/r_{1}^{(\rm no)}) r_2^{(\rm no)}}{r_2^{(\rm td)}}, t_{1}^{(\rm c)}\r\},\label{p4case2t2}\\
t^{(\rm no)*} &= (L_2-t_{2}^{(\rm td)*}r_2^{(\rm td)})/r_{2}^{(\rm no)},\\
 t_{1}^{(\rm td)*}& = (L_1-t^{(\rm no)*}r_{1}^{(\rm no)} )/r_1^{(\rm td)},\label{p4case2t1}\
 \end{align}
 which leads to $T_2^*=t_{1}^{(\rm td)*}+t^{(\rm no)*} +t_{2}^{(\rm td)*} + t_{2}^{(\rm c)}$.
  }
  \end{itemize} 
}
\end{proposition}
\noindent\emph{Sketch of proof:}
First, if $t_{1}^{(\rm c)}\ge L_2/r_2^{(\rm td)}$, we have $\max(0,t_{1}^{(\rm c)}-t_{2}^{(\rm td)})=t_{1}^{(\rm c)}-t_{2}^{(\rm td)}$ and hence the objective function of problem (P$4$) is equivalent to $t_{1}^{(\rm td)}+t^{(\rm no)}+t_{1}^{(\rm c)}+t_{2}^{(\rm c)}$, whose optimal solution can be easily obtained in \eqref{p4case1}. On the other hand, if $t_{1}^{(\rm c)}< L_2/r_2^{(\rm td)}$, we first consider an auxiliary problem that neglects the term of $\max(0,t_{1}^{(\rm c)}-t_{2}^{(\rm td)})$ in the objective function of problem (P$4$), for which the optimal solution can be obtained as in Proposition~\ref{prop:infinite}, denoted by $\{\tilde{t}_{1}^{(\rm td)}, \tilde{t}_{2}^{(\rm td)}, \tilde{t}^{(\rm no)}\}$. Then, if $\tilde{t}_{2}^{(\rm td)}\ge t_1^{(\rm c)}$ (i.e., $L_2/r_2^{(\rm td)}-L_1r_2^{(\rm no)}/(r_{1}^{(\rm no)} r_2^{(\rm td)})\ge t_{1}^{(\rm c)}$), we have $\max(0,t_{1}^{(\rm c)}-\tilde{t}_{2}^{(\rm td)})=0$ and hence $\{\tilde{t}_{1}^{(\rm td)}, \tilde{t}_{2}^{(\rm td)}, \tilde{t}^{(\rm no)}\}$ is still the optimal solution to problem (P$4$). Otherwise, it can be proved by contradiction that the optimal $t_{2}^{(\rm td)*}$ should be $t_{2}^{(\rm td)*}=t_1^{(\rm c)}$ that leads to $\max(0,t_{1}^{(\rm c)}-t_{2}^{(\rm td)*})=0$ and the corresponding $t^{(\rm no)*}$ and $ t_{1}^{(\rm td)*}$ can be obtained in \eqref{p4case2t2}--\eqref{p4case2t1}. Combining the above leads to the results in Proposition~\ref{prop2}.
\hfill $\Box$

Proposition~\ref{prop2} shows that the task-computing time of the firstly-scheduled user significantly affects the transmission time division and scheduling. Specifically, if user $1$ incurs a relatively long computing delay (i.e., $t_{1}^{(\rm c)}\ge L_2/r_2^{(\rm td)}$), TDMA-based data transmission is preferred owing to the increased transmission rates of both users as well as the maximum time overlapping between user $1$'s task computing and user $2$'s data transmission. Otherwise, the optimal offloading scheduling is determined by the defined NOMA-priority, and NOMA and TDMA are preferred when $\lambda({\boldsymbol \theta}^{(\rm no)}, {\bf p})$ is above and below a threshold (i.e., $0$), respectively. Moreover, it is worth noting that even for the case of  $t_{1}^{(\rm c)} < L_2/r_2^{(\rm td)}$ and $\lambda({\boldsymbol \theta}^{(\rm no)}, {\bf p})\ge 0$, the optimal NOMA-based transmission with finite cloud capacity may be different from that of the infinite-cloud-capacity case. To be specific, for the former case, if $(L_2-(L_1/r_{1}^{(\rm no)}) r_2^{(\rm no)})/r_2^{(\rm td)}< t_{1}^{(\rm c)}$, user $2$ should make full use of the task-computing interval for user $1$ to transmit part of its data for reducing the sum delay since it does not incur addition delay.

Next, similarly to the case with infinite cloud capacity, it can be shown that the optimal solution to problem (P$1$) for the  finite-cloud-capacity case can be numerically obtained by searching over all possible combinations of IRS phase shifts with $\{p_1=P_1, p_2=P_2\}$ and choosing the one that achieves the minimum sum delay as given in Proposition~\ref{prop2}.
\section{Numerical Results}
\begin{figure}[t]
 \centering
  \includegraphics[width=3.5in]{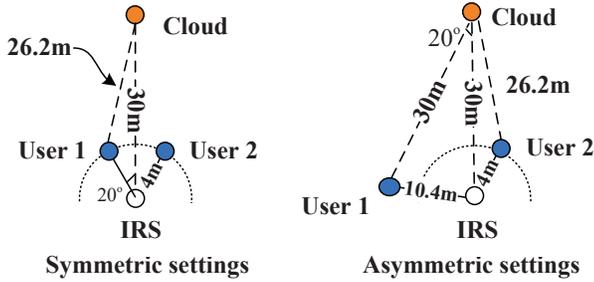}
 \caption{Symmetric and asymmetric user deployment.}
  \label{top}
\end{figure}
Numerical results are presented in this section to demonstrate the effectiveness of the proposed time-sharing NOMA  scheme in the IRS-aided MEC system. Without otherwise specified, we consider an IRS with $N=5$ subsurfaces, each consisting of $M=20$ reflecting elements.  For the large-scale path loss, the channel power gain at the reference distance of $1$ meter (m) is set as $-30$ dB, and the path loss exponents for the channels from user $k$ to AP, user $k$ to IRS, and IRS to  AP are set as $3.2$, $2.6$, and $2.6$, respectively. The small-scaling fading for the involved channels is modeled by the Rayleigh fading. We consider two user-deployment cases as shown in Fig.~\ref{top}. Other parameters are set as $P_1=P_2= 5$ dBm, $B=250$ KHz, $\sigma^2=-140$ dBm/Hz, $C_1 = C_2 = 300$ cycles/bit, and $F = 5$ GHz (if not specified). 

\begin{figure}[t]
\centering
\includegraphics[width=8.5cm]{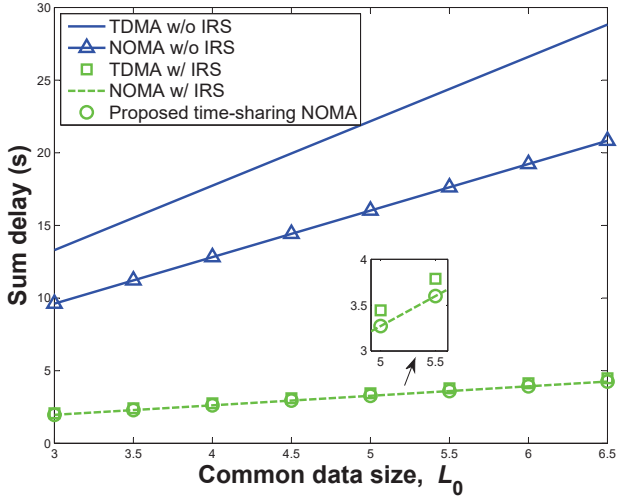}
\caption{Sum delay versus the common data size $L_0$ under symmetric user deployment.}
\label{SP1}
\end{figure}

\begin{figure}[t]
\centering
\includegraphics[width=8.5cm]{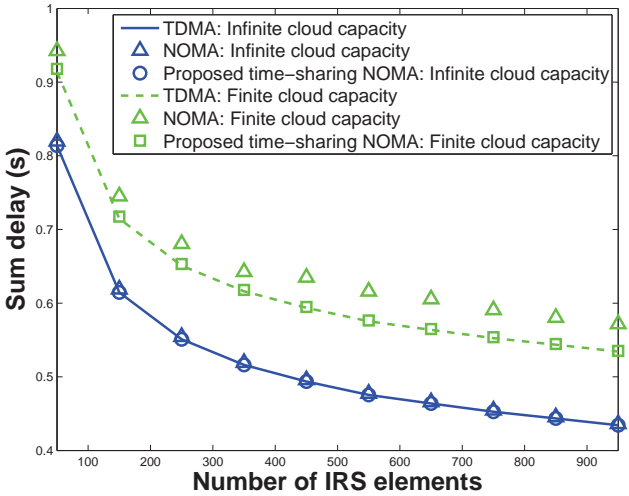}
\caption{Sum delay versus the number of IRS elements under symmetric user deployment.}
\label{SP4}
\end{figure}

We first consider the symmetric user-deployment case in Fig.~\ref{top}, where the two users have equal distances from the cloud as well as from the IRS. Fig.~\ref{SP1} shows the sum delay versus the common data size (i.e., $L_1=L_2=L_0$) of the proposed time-sharing NOMA  scheme against two benchmark schemes assuming infinite cloud capacity: 
1) TDMA: the two users are scheduled based on TDMA with $t^{(\rm no)}=0$; 2) NOMA: the two users firstly transmit data based on NOMA and then one of them sends its remaining data alone (i.e., $t^{(\rm no)}\neq0$ and $t_{1}^{(\rm td)}(\text{or}~ t_{2}^{(\rm td)})=0$). Both the systems with/without IRS are considered. First, it is observed that all the schemes with IRS outperform their counterparts without IRS owing to the passive beamforming gain. Next, for the system with IRS, the proposed time-sharing NOMA scheme has similar sum delay with the NOMA benchmark, which is smaller  than that of TDMA. This is because with symmetric deployment and infinite cloud capacity, the {NOMA priority} is likely positive  and thus the proposed scheme reduces to NOMA. Moreover, different from the case without IRS for which NOMA always achieves much smaller sum delay than TDMA, the performance gap between the NOMA and TDMA benchmarks with IRS is much smaller.
In Fig.~\ref{SP4}, we further show the sum delay versus the number of IRS elements under finite/infinite cloud capacities  with $N = 50$, where $L_1 = L_2 = 1$ Mbits and $C_1 = C_2 = 500$ cycles/bit. One can observe that the sum delay for all schemes monotonically decreases with the number of IRS elements. In addition, given infinite cloud capacity, TDMA slightly outperforms NOMA while the proposed scheme reduces to TDMA and both achieve the smallest sum delay. The sum-delay reduction of TDMA as compared to NOMA  is more pronounced under the finite cloud capacity, due to  the  transmission-computing multiplexing gain of the TDMA-based scheduling.

\begin{figure}[t]
\centering
\includegraphics[width=8.5cm]{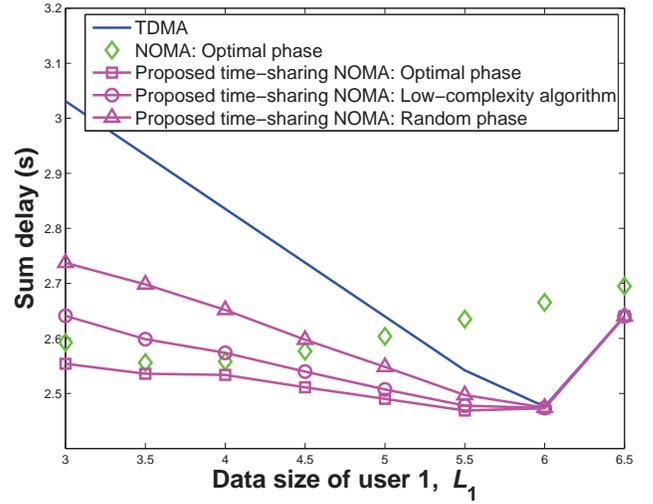}
\caption{Sum delay versus the data size of near-IRS user $1$ with $L_1+L_2=6.7$ Mbits under asymmetric user deployment.}
\label{SP2}
\end{figure}

\begin{figure}[t]
\centering
\includegraphics[width=8.5cm]{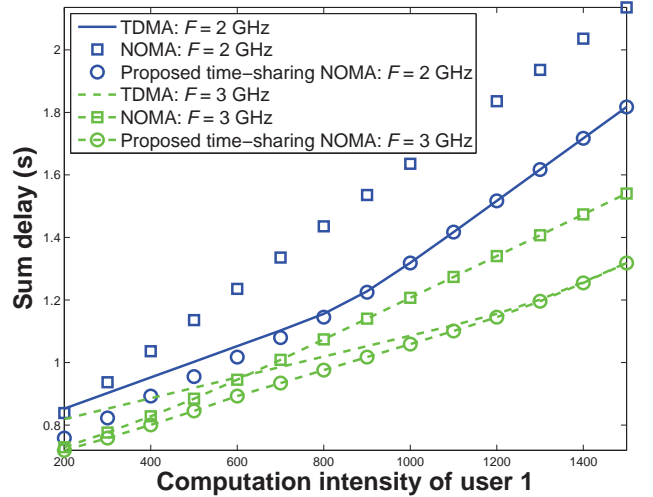}
\caption{Sum delay versus the computation intensity of user $1$ under asymmetric user deployment.}
\label{SP3}
\end{figure}

Next, we consider the  asymmetric user-deployment case in Fig.~\ref{top} with one near-IRS user and one far-IRS user, indexed by users $1$ and $2$, respectively. In Fig.~\ref{SP2}, we compare the sum delay versus the data size of user $1$ under the proposed time-sharing NOMA scheme with the following three phase-shift designs against the TDMA and NOMA benchmarks: 1) optimal phase based on the exhaustive search; 2) low-complexity algorithm based on the proposed phase linearization; 3) random phase for which the IRS randomly generates $5$ sets of phase shifts and selects the best one that achieves the minimum sum delay. %The system parameters are set as $C_1=C_2=300$ cycles/bit, $F=5$ GHz, 
The sum data size of the two users is fixed as $L_1+L_2=6.7$ Mbits. It is observed that as $L_1$ increases, the sum delay of the proposed time-sharing NOMA scheme firstly decreases and then linearly increases when $L_1\ge 6$ Mbits. This is because for small $L_1$, increasing $L_1$ leads to a smaller sum transmission delay since user $1$ is nearby the IRS and thus has a smaller path loss; while when $L_1\ge 6$ Mbits, the proposed time-sharing NOMA scheme reduces to the TDMA benchmark for which the sum delay linearly increases with $L_1$ (see Proposition~\ref{prop2}). Moreover, the proposed scheme based on the low-complexity algorithm outperforms that based on the random phase, but suffers some performance loss as compared to that based on the optimal phase design. Last, we show in Fig.~\ref{SP3} the effects of computation intensity of user $1$ (i.e., $C_1$) on the sum delay with $L_1=L_2=1$ Mbits. It is observed that the sum delay for all schemes monotonically increases with $C_1$. Moreover, the proposed time-sharing NOMA scheme outperforms and achieves the same delay performance with the TDMA benchmark when $C_1$ is below and above a certain threshold (i.e., ${L_2 F}/{L_1 r_2^{(\rm td)}}$), respectively, since the proposed time-sharing NOMA scheme reduces to TDMA when $C_1$ is sufficiently large.

\section{Conclusions}
In this paper, we proposed a flexible time-sharing NOMA scheme for an IRS-aided two-user MEC system that encapsulates both conventional NOMA and TDMA as special cases. We formulated and solved an optimization problem for minimizing the sum delay of the two users under the IRS discrete-phase constraint.
In particular, we analytically showed that TDMA is better than NOMA for the sum delay minimization when the users' cloud-computing time is sufficiently long and/or the rate improvement of TDMA against NOMA, both  with IRS, is sufficiently large; whereas the relationship is reversed otherwise.
Numerical results were presented to corroborate our main findings.
\bibliographystyle{IEEEtran} 
\bibliography{MEC_IRS_CITE}
\end{document}